\documentclass[aps,prl,twocolumn,superscriptaddress,preprintnumbers,floatfix,nofootinbib]{revtex4}
\usepackage{graphicx}
\usepackage{epsfig}

\begin{document}

\newcommand{\be}{\begin{eqnarray}}
\newcommand{\ee}{\end{eqnarray}}
\newcommand\del{\partial}
\newcommand{\nn}{\nonumber } 
\newcommand{\tK}{Z_{n=1}}
\newcommand{\re}{\mathrm{Re}}
\newcommand{\im}{\mathrm{Im}}


\title  {Lattice simulations of QCD with $\mu_B\neq0$ versus phase
  quenched QCD}

\author{K. Splittorff}
\affiliation{Nordita, Blegdamsvej 17, DK-2100, Copenhagen {\O}, Denmark}

\date   {\today}

\begin  {abstract}
Previously published lattice results for QCD at $\mu_B\neq0$ are compared 
to analytic predictions for phase quenched QCD. We observe that the strength
of the sign problem in QCD is linked directly to the position of the
phase transition line for pion condensation in phase quenched QCD and that
the number of terms needed in the Taylor expansion approach depends on the
strength of the sign problem. 
Together this emphasizes the physical importance of the sign problem and helps 
to clarify the range over which the Taylor expansion approach is 
practically applicable. 
Finally, we observe that the positions of the endpoint of the first order
chiral phase transition in the QCD
phase diagram found in two successive computations by Fodor and Katz are both 
close to the position of the phase transition line for pion condensation
in phase quenched QCD.     
\end {abstract}

\maketitle

\section{Introduction} 

Results from lattice simulations for unquenched QCD at non-zero baryon 
chemical potential give us valuable information about the
non-perturbative sector of strongly interacting matter. 
The central aim of the lattice simulations at non-zero 
baryon chemical potential is to determine the phase diagram of unquenched 
QCD. 
The focus at present is on the chiral transition as a function of temperature
and rather small chemical potential. 
Predictions for the slope of the chiral crossover line at zero chemical
potential and for the position of the endpoint of the first order chiral
phase transition have already been set forth. 
Lattice simulations at non-zero baryon chemical potential are, however,  
different from lattice simulations at zero baryon chemical potential in that
they have to deal with the sign problem. 
The term 'sign problem' is used to describe the numerical
difficulties in doing Monte Carlo sampling on a non-positive weight. 
The presence of the sign problem is a direct consequence of the
imbalance between quarks and anti-quarks one imposes in order to generate  
a non-zero baryon density. Moreover, the sign problem has several direct
physical implications. One of the most dramatic can be seen if we compare to 
phase quenched QCD. 

{\sl Phase quenched QCD} is different from QCD in that one takes the
absolute value of the fermion determinant in the measure
\be
Z_{\rm PQ} = \int [{\rm d} A_\eta]
|\det(D_\eta\gamma_\eta+\mu\gamma_0+m)|^{N_f}e^{-S_{\rm YM }}.
\ee
The phase diagram of this theory has an extended region which is dominated by
a Bose-Einstein condensate of pions. Returning to QCD by including the phase 
of the determinant wipes out this pion phase entirely. 
In order to deal with the phase of the determinant in lattice simulations of
QCD  three main approaches have been pursued: 1) the Taylor expansion approach
\cite{Bielefeld-Swansea1,Bielefeld-Swansea2,TaylorGupta,GG,Allton6}, 2) the
reweighting approach \cite{Glasgow,Fodor:2001au,FK1,FK2,Bielefeld-Swansea1}, 
and 3) the imaginary
chemical potential approach \cite{Philipsen1,Philipsen2,ImagmuLombardo}.

In this paper we address some of the issues involved in the interpretation
of existing lattice data for QCD at non-zero baryon chemical potential 
obtained with these approaches. We will do so by comparing previously 
published lattice data for QCD at non-zero baryon chemical potential 
to phase quenched QCD. First of all we replot 
a lattice measurement from \cite{Allton6} of the variance of the phase of the
determinant in units where we can compare to analytic predictions for phase
quenched QCD. We observe that the contour lines for the variance of the
phase of the determinant are aligned with the phase transition line for pion
condensation in phase quenched QCD. Moreover, the distance between the contour
lines decreases as the values of $\mu$ and $T$ approach the region where phase
quenched QCD is in the pion phase. Next, we wish to verify, as stated in
\cite{GG}, that the order of the Taylor expansion needed depends on the 
strength of the sign problem. In order to do so we consider the 6th order
Taylor expansion of the quark number susceptibility published in
\cite{Allton6}. In agreement with the statement of \cite{GG} we 
find that the 6th order term in the expansion becomes important when the
variance of the phase exceeds a certain value. Together with the first
observation we conclude that it will be exceedingly hard to make predictions
for QCD at non-zero baryon chemical potential using the Taylor expansion
approach when the values of $\mu$ and $T$ are such that pion condensation
occurs in phase quenched QCD. 

Finally, we consider the values of $\mu$ and $T$ at the endpoint of the first
order chiral phase transition as found
using the reweighting method in \cite{FK1} and \cite{FK2}. After rescaling the 
coordinates for the endpoint we observe that both predictions occur at values
of $\mu$ and $T$ which are close to the phase boundary for pion
condensation in phase quenched QCD.

\section{Pion condensation in phase quenched QCD}

At zero temperature the critical chemical potential for pion condensation in
phase quenched QCD is $\mu=m_\pi/2$ and at small temperature the critical 
chemical potential can be evaluated in chiral perturbation theory
\cite{STV2}. The critical temperature found at 1-loop order corresponds to 
the semi-classical result for the critical temperature of Bose-Einstein 
condensation in a dilute massive Bose gas. The general determination of 
this phase transition can be evaluated directly by means of Monte Carlo
simulations of phase quenched QCD and such 
studies are currently taking place \cite{Kogut}. One can also estimate 
the position of the phase transition using phenomenological models.     
Probably the simplest prediction for the critical temperature in phase quenched QCD
is obtained from the random matrix model considered in \cite{KTV}. 
This mean field approach leads to a critical temperature (see eq
(5.48) of \cite{KTV}) which depends both on the chemical potential, 
the quark mass and the phenomenological constant introduced in the model. 
However, by taking 
the quark mass to zero while keeping the ratio of the pion mass to the 
chemical potential fixed the dimensionful constant drops out and the 
result for the critical temperature for pion condensation in phase 
quenched QCD is ($T_0$ is the pseudo critical temperature for chiral 
symmetry breaking at $\mu=0$)  
\be
T_c/T_0 = \sqrt{1-\left(\frac{m_\pi}{2\mu}\right)^4}.
\label{Tcmu}
\ee
In the range $m_\pi/2<\mu< m_\pi$ this simple result agrees well 
with the predictions from the NJL model \cite{HMZ,VJ}, the prediction from 
strong coupling QCD \cite{Nishida}, as well as the predictions from a random 
matrix model with all Matsubara frequencies included \cite{VJ}. At larger 
values of the chemical potential the critical temperature obtained in these 
models drops down to zero again as a result of saturation.

The $\mu$ dependence of the critical temperature given in (\ref{Tcmu}) is 
also consistent with the picture emerging from the lattice studies of phase
quenched QCD \cite{Kogut}. These lattice simulations and analytical arguments
\cite{STV2,SLV} however suggest that the order of the pion phase
transition changes from 2nd order to 1st order with increasing chemical
potential. This aspect is not reproduced by the random matrix model.

\section{The variance of the phase}

\begin{figure}[ht]
  \unitlength1.0cm
  \begin{center}
  \begin{picture}(3.0,2.0)
  \put(-3.,-6.){
    \epsfig{file=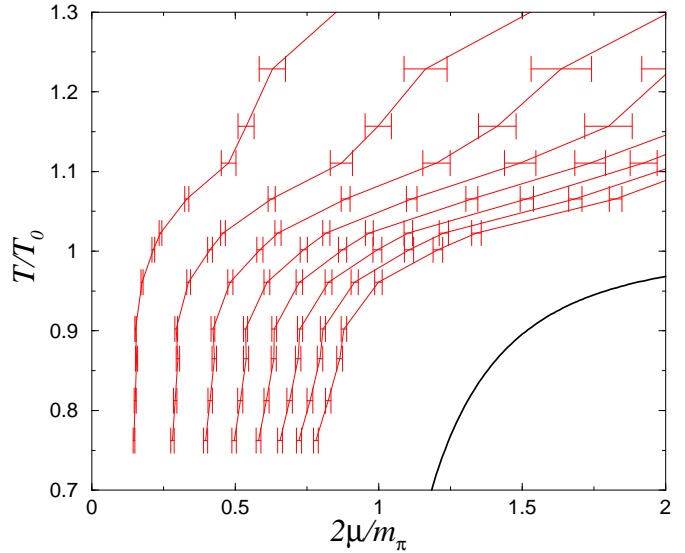,clip=,width=8.8cm}}
  \end{picture}
  \vspace{5.5cm}
  \caption{ 
  \label{fig:1}
In the ($2 \mu/m_\pi$,$T/T_0$)-plane we plot \\
$\bullet$ In red: Lines of constant variance $\sqrt{\langle\theta^2\rangle-\langle\theta\rangle^2}$ from
\cite{Allton6}. From the upper left the values increase from $\pi/4$ to
$2\pi$ in steps of $\pi/4$. Lines of higher variance are not
available. \\ 
$\bullet$ In black: The critical temperature as a function of the chemical
potential for pion condensation in phase quenched QCD as estimated in
(\ref{Tcmu}).}
  \vspace{-6mm}
  \end{center}
\end{figure}

A direct measure of the sign problem in QCD is the variance of the phase,
$\theta$, of the fermion determinant. The phase was measured in 
\cite{Shinji,Allton6} and a contour diagram with lines of constant variance,
$\sqrt{\langle \theta^2 \rangle-\langle \theta \rangle^2}$, in the 
$(T/T_0,\mu/T)$-plane was given. Below we re-plot the data from 
\cite{Allton6} as a contour diagram in the $(2\mu/m_\pi,T/T_0)$-plane,
$T_0$ again being the pseudo-critical temperature for the crossover 
at $\mu=0$.   
The value of the pion mass used in order to convert the plot from \cite{Allton6} 
is $m_\pi/T_0=3.58$. Shown are contour lines for $\sqrt{\langle \theta^2 \rangle-\langle \theta \rangle^2}
=\pi/4,\pi/2, 3\pi/4,\ldots,2\pi$, the value increasing with the chemical 
potential. Note that contour lines for  $\sqrt{\langle \theta^2
  \rangle-\langle \theta \rangle^2}>2\pi$ are not displayed. 
In addition to the contour lines the prediction (\ref{Tcmu}) 
for the critical temperature for pion condensation in phase quenched QCD is 
shown. 
The contour lines are seen to be aligned with the phase transition line and 
the distance between the contour lines decreases as the phase transition line 
is approached. This observation is consistent with the fact that the 
presence of the phase of the determinant wipes out the pion phase transition.
In particular, it is natural to expect that 
$\sqrt{\langle \theta^2 \rangle-\langle \theta \rangle^2} >> 1$ in the region
where phase quenched QCD enters the pion phase, even with a moderate
volume. 
At zero temperature this has been verified analytically 
\cite{O,AOSV}; the eigenvalue density of the Dirac operator in QCD with
$\mu\neq0$ becomes highly oscillating for $\mu>m_\pi/2$ and it is these 
oscillations which insure that the pion phase is avoided in QCD \cite{OSV}. 
Note that the contour lines shown in figure \ref{fig:1} will shift toward 
lower values of $\mu$ and higher values of $T$ with increasing volume (for a
discussion of the volume dependence of the variance see \cite{Shinji}).    

\section{Taylor expansions} 

One of the approaches to QCD at non-zero quark number chemical potential which
are being pursued 
\cite{Bielefeld-Swansea2,TaylorGupta,GG,Allton6} 
is based on a Taylor expansion in $\mu/T$ around $\mu=0$. 
First, the forms of the traces contributing to an operator at a given order
in the chemical potential are determined analytically and then these traces are
evaluated by standard Monte Carlo methods at $\mu=0$.     
Traces contributing to order $2n$ in the Taylor expansion of an
extensive observable are typically of order $V^n$ and these terms must 
combine to a result linear in the volume \cite{GG}.  
This makes it difficult to control numerical errors in a high order
expansion. 

\begin{figure}[t]
  \unitlength1.0cm
  \begin{center}
  \begin{picture}(3.0,2.0)
  \put(-3.,-5.5){
    \epsfig{file=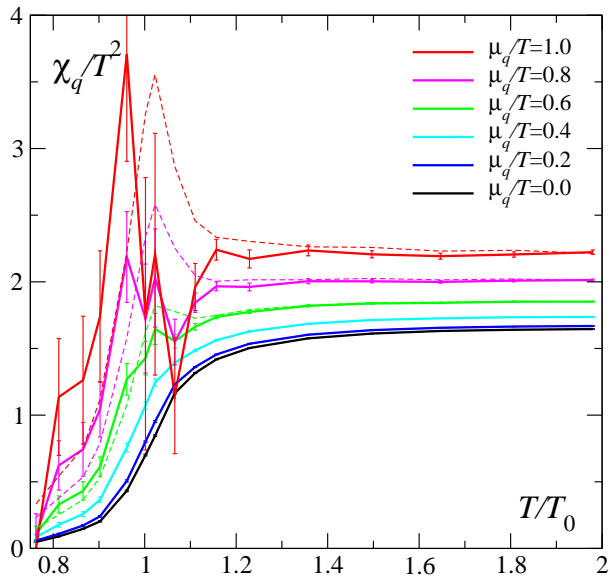,clip=,width=8cm}}
  \end{picture}
  \vspace{5.5cm}
  \caption{ 
  \label{fig:2}
Figure from \cite{Allton6} showing the quark number susceptibility for fixed
values of $\mu/T$ as a function of temperature. Full lines give the result 
of the 6th order Taylor expansion while the dashed lines give the results at 
4th order.}
  \vspace{-6mm}
  \end{center}
\end{figure}

\begin{figure}[t]
  \unitlength1.0cm
  \begin{center}
  \begin{picture}(3.0,2.0)
  \put(-3.,-5.5){
    \epsfig{file=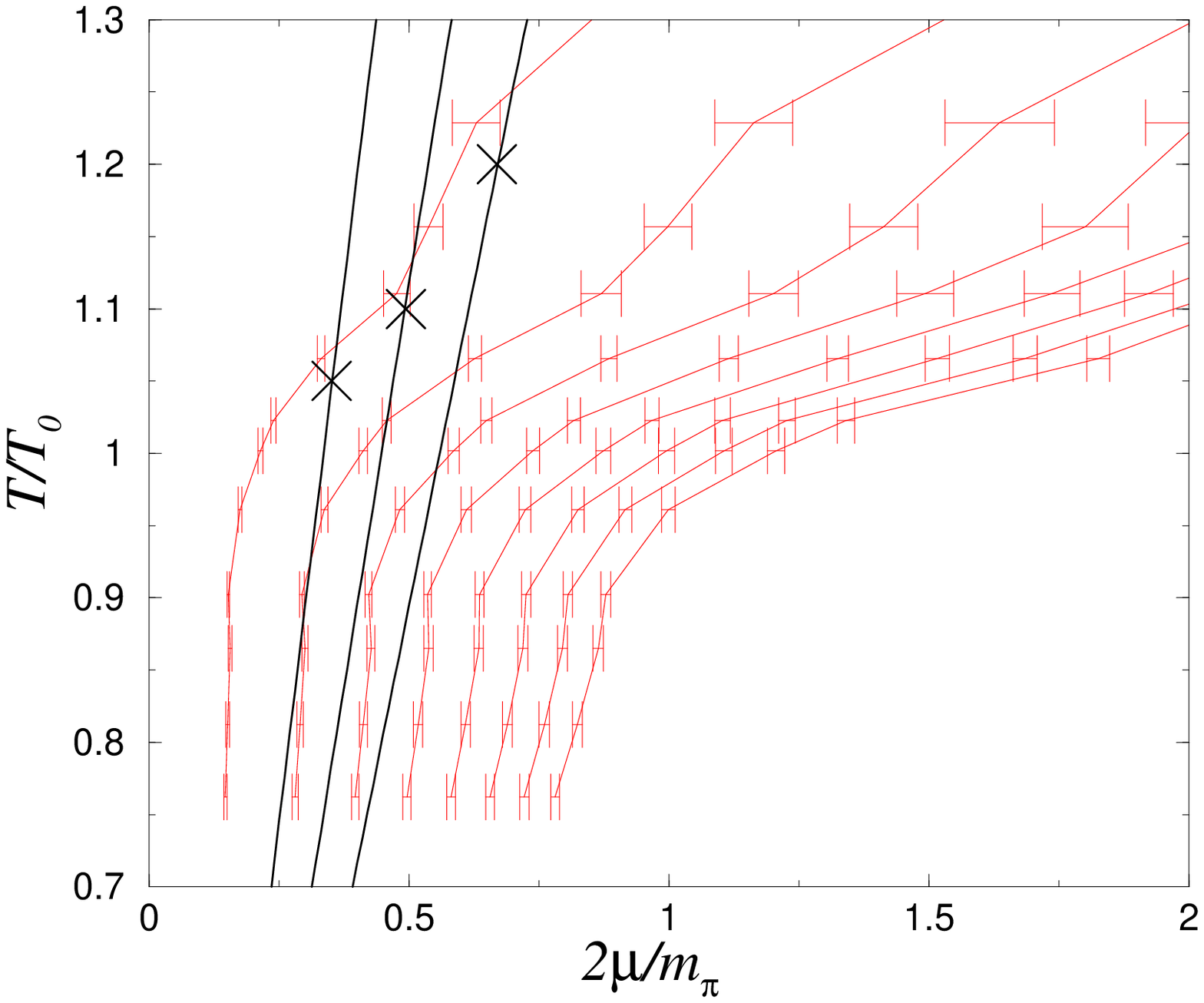,clip=,width=8.8cm}}
  \end{picture}
  \vspace{5.5cm}
  \caption{ 
  \label{fig:3} 
In the ($2 \mu/m_\pi$,$T/T_0$)-plane we plot\\
$\bullet$ Red lines:
Lines of constant $\sqrt{\langle\theta^2\rangle-\langle\theta\rangle^2}$ from
\cite{Allton6}. From the upper left the values increase from $\pi/4$ to
$2\pi$ in steps of $\pi/4$.
\\
$\bullet$ Black lines:
Lines with fixed ratio of $\mu/T$, the values shown are $\mu/T=0.6$, 0.8, and
1.0.  \\
$\bullet$ Black crosses: Marks the point on the black lines below which the
6th order term in the Taylor expansion of the quark number susceptibility in 
\cite{Allton6} is important.}
  \vspace{-6mm}
  \end{center}
\end{figure}

In \cite{GG} it is stated that the order of terms needed in the Taylor
expansion depends on the severity of the sign problem. 
Here we verify this statement using the quark number susceptibility
determined in \cite{Allton6} to 6th order in $\mu$. The relevant figure 
from \cite{Allton6} is given in figure \ref{fig:2} for convenience. 
The quark number susceptibility is shown as a function
of temperature for fixed values of $\mu/T$ as obtained from the 6th order
Taylor expansion. For the three largest values of $\mu/T$ the result from the
4th order Taylor expansion are indicated by the dashed lines. From this plot
we estimate the values of $T/T_0$ below which the 6th order term is
important. Using simply the magnitude of the deviation in the units of figure
\ref{fig:2} for $\mu/T=1.0$ we find $T/T_0\simeq1.20$, 
for  $\mu/T=0.8$ we find $T/T_0\simeq1.10$, and for 
$\mu/T=0.6$ we find $T/T_0\simeq1.05$. In figure \ref{fig:3} we plot 
the contour lines for the variance of $\theta$ from \cite{Allton6} in the
$(2\mu/m_\pi,T/T_0)$ plane, this time together with the 3 lines determined by
$\mu/T=1.0$, 0.8, and 0.6.     
On the lines of constant $\mu/T$ we have just estimated the values of $T/T_0$
below which the 6th order term in the Taylor expansion of the quark number
susceptibility becomes important. These values are indicated by the three crosses 
in figure \ref{fig:3}. We observe that the 6th order term in the 
determination of the quark number susceptibility becomes important for values
of $T$ and $\mu$ where the variance of the phase is larger than $\pi/4$. 
That is, the 4th order Taylor expansion of the quark number susceptibility in 
\cite{Allton6} is only appropriate to the left of the leftmost contour 
line in \ref{fig:2}. 
Given the sufficient numerical accuracy to handle the
delicate cancellations of the terms of order $V^6$, the 6th order
Taylor expansion of the quark number susceptibility approach may be adequate 
somewhat to the right of the
$\sqrt{\langle\theta^2\rangle-\langle\theta\rangle^2}=\pi/4$ contour line. 

Combined with the observations made from figure \ref{fig:1} we conclude that
even for small volumes it will be extremely difficult to make predictions for 
observables in QCD using the Taylor expansion approach when the values of 
$\mu$ and $T$ are such that phase quenched QCD is in the pion phase.  

Certainly, our estimate above only serves as a first numerical test of 
the proposition that the number of terms needed in the Taylor expansion 
depends on the strength of the sign problem. A dedicated series of
measurements of several observables on the lattice are needed in order to
settle this firmly.

\section{Critical endpoint}

\begin{figure}[t]
  \unitlength1.0cm
  \begin{center}
  \begin{picture}(3.0,2.0)
  \put(-3.,-5.5){
    \epsfig{file=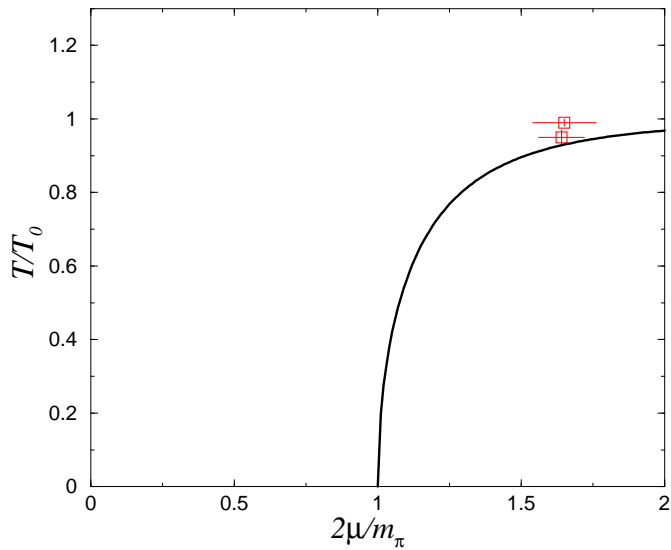,clip=,width=8.8cm}}
  \end{picture}
  \vspace{5.5cm}
  \caption{ 
  \label{fig:4} 
In the ($2 \mu/m_\pi$,$T/T_0$)-plane we plot\\
$\bullet$ Red squares:
The position of the critical endpoint as computed in \cite{FK1} (lower point) and in \cite{FK2} (upper point). 
\\
$\bullet$ Black line:
 The critical temperature as a function of the chemical potential for pion
 condensation in phase quenched QCD as estimated in (\ref{Tcmu}).  
}
  \vspace{-6mm}
  \end{center}
\end{figure}

In this section we compare the positions of the endpoints of the first order
chiral phase transition as determined in \cite{FK1} and \cite{FK2} to the 
estimated position of the pion phase transition in the phase quenched 
theory\footnote{I thank Misha 
Stephanov for suggesting this comparison.}. In order to facilitate
this comparison we again consider the plane $(2\mu/m_\pi,T/T_0)$ and plot in
figure \ref{fig:4} simultaneously the critical endpoints determined for QCD in 
\cite{FK1,FK2} and the position of the pion phase transition 
for phase quenched QCD given in (\ref{Tcmu}). The values of the pion mass 
used in order to make the rescaling was $m_\pi=294$MeV for \cite{FK1} 
and $m_\pi=145$MeV for \cite{FK2}.  

We observe that the endpoints found in
\cite{FK1} and \cite{FK2} both are remarkably close to the phase 
transition line estimated in (\ref{Tcmu}) for phase quenched QCD. 
(The fact that the chemical potential at the endpoint scales with $m_\pi$ was
noted in \cite{GG}.)

The phase transition into the pion phase occurring in phase quenched QCD 
has a very simple manifestation in the eigenvalue spectrum of the Dirac 
operator: it occurs when the density of eigenvalues of the Dirac operator reaches
the quark mass (see e.g. \cite{diracPQnum} or \cite{diracPQana,AOSV}). That
is, the critical chemical potential and temperature are determined by 
\be
\rho_{\rm PQ}(z=m,m,T_c,\mu_c) \neq 0, 
\ee
where the eigenvalues are given by
\be
(D_\eta\gamma_\eta+\mu\gamma_0)\psi_n =z_n\psi_n
\ee
and the eigenvalue density is 
\be
\rho_{\rm
  PQ}(z,m,T,\mu)=\left\langle\frac{1}{V}\sum_{n}\delta^2(z_n-z)\right\rangle_{\rm PQ}.
\ee
For a lattice simulation involving the full determinant of the QCD Dirac operator 
this leads to substantial problems: For $\mu$ larger than the critical $\mu$
for pion condensation in phase quenched QCD the eigenvalues $z_n$ of 
$D_\eta\gamma_\eta+\mu\gamma_0$ lie arbitrarily close to $-m$. The complex phase of
$(z_n+m)$ therefore is extremely sensitive to the gauge field configuration 
and this manifests itself in the fluctuation of the phase of the fermion 
determinant. For this reason 
the reweighting methods are exceedingly delicate for values of $\mu$ and $T$
where phase quenched QCD is in the pion phase. 
In the light of this, it is unfortunate that the predictions for the critical
endpoint in both \cite{FK1} and \cite{FK2} are so close to the predicted 
phase transition for phase quenched QCD.

\section{Conclusions} 

In order to emphasize the physical importance of the sign problem for present lattice 
calculations at non-zero baryon chemical potential we have replotted 
previously published lattice data in units where a direct comparison to phase 
quenched QCD is possible. We have observed that the seriousness of the 
sign problem, as measured by the variance of the phase of the fermion
determinant, depends on the distance in the ($2\mu/m_\pi,T/T_0$)-plane from
pion phase transition in phase quenched QCD. 
The estimate used for the position of the phase transition in phase quenched QCD 
is independent of any phenomenological parameters. It is obtained as the 
limit of small quark masses with $\mu/m_\pi$-fixed of the result from a 
random matrix model. While this result is in reasonable agreement with
other phenomenological models one should only take it as a rough guideline
for the position of the pion phase transition in phase quenched QCD. 
Lattice simulations of phase quenched QCD can measure the position as well as the 
order of the phase transition and thus allow for a direct comparison between
QCD and phase quenched QCD.

Using the result for the Taylor expansion of the quark number susceptibility
in \cite{Allton6} we have verified that the number of terms needed in the
Taylor expansion approach depends on the seriousness of the sign problem as measured
by the variance of the phase of the fermion determinant. The need for a
Taylor expansion to $2n$th order post a demand for sufficient numerical accuracy to
realize cancellations between terms of order $V^n$ in order to obtain an
expectation value of order $V$ \cite{GG}. Thus the sign problem sets the
practical upper bound on the applicability of the Taylor expansion approach.    
In particular it will be extremely computationally demanding to determine
observables in QCD by means of a Taylor expansion if the values of $\mu$ and
$T$ are such that phase quenched QCD is in the pion phase. 
   
The need for a high order Taylor expansion to penetrate regions of the
($\mu$,$T$)-plane where the sign problem is strong also sets a practical 
limit on the imaginary chemical potential approach. The analytic continuation from 
imaginary to real chemical potential is carried out by fitting a
polynomial in $\mu$ and the order of the polynomial needed corresponds to 
that of the Taylor expansion. We have checked that no data points have been
reported in the literature with this method for values of $\mu$ and $T$ where
phase quenched QCD is in the pion condensed phase.   

Finally, we have observed that the location of the endpoint of the first order
chiral phase transition determined using the 
reweighting method in \cite{FK1} as well as in \cite{FK2} falls surprisingly
close to the location of the pion phase transition in the phase quenched
theory. Despite attempts, no analytical argument to date suggests that
this should be the case. For values of $T$ and $\mu$ where the phase quenched
theory enters the pion phase the reweighting approach is expected to be 
extremely delicate. Hence, one may worry that the critical point in 
\cite{FK1} as well as in \cite{FK2} are manifestations of the numerical
difficulties encountered rather than a true physical effects. 
Certainly it would be interesting if by an appropriate choice of quark masses
one could separate the prediction for the critical point in QCD from the
phase boundary in phase quenched QCD.  As suggested in \cite{Philipsen2} fine
tuning the quark masses close to critical value for which the chiral
transition at $\mu=0$ becomes first order may cause the endpoint to move to 
smaller values of $2\mu/m_\pi$.  

\vspace{1mm}  

\noindent
{\bf Acknowledgments:} It is a pleasure to thank the participants in and
organizers of the KITP program {\sl Modern challenges for lattice 
gauge theory} for many stimulating discussions and clarifying answers. In
particular Maria Paula Lombardo, Dominique Toublan, Misha Stephanov, Owe
Philipsen, Philippe de Forcrand, Jac Verbaarschot and Don Sinclair are thanked. In addition 
I am grateful to Simon Hands and Shinji Ejiri for challenging discussions and 
sharing data files.


\begin{thebibliography}{9}



\bibitem{Bielefeld-Swansea1}
C.R. Allton, S. Ejiri, S.J. Hands, O. Kaczmarek, F. Karsch, E. Laermann, 
 C. Schmidt and L. Scorzato, Phys. Rev. D {\bf 66} (2002) 074507.

\bibitem{Bielefeld-Swansea2}
C.R.~Allton, S.~Ejiri, S.J.~Hands, O.~Kaczmarek, F.~Karsch, E.~Laermann and
 C.~Schmidt,
 Phys.\ Rev.\ D {\bf 68} (2003) 014507.

\bibitem{TaylorGupta}
R.V. Gavai and S. Gupta, Phys. Rev. D {\bf 64} (2001) 074506;
 R.V. Gavai and S. Gupta, Phys. Rev. D {\bf 68} (2003) 034506;
 R.V. Gavai, S. Gupta and R. Roy, Prog.\ Theor.\ Phys.\ Suppl.\  {\bf
 153} (2004) 270; 
R.V. Gavai, S. Gupta and P. Majumdar, Phys. Rev. D {\bf 65} (2002) 054506.

\bibitem{GG}
R.V. Gavai and S. Gupta, 
hep-lat/0412035. 

\bibitem{Allton6}
  C.~R.~Allton {\it et al.},
  Phys.\ Rev.\ D {\bf 71} (2005) 054508.


\bibitem{Glasgow} 
  I.~M.~Barbour, S.~E.~Morrison, E.~G.~Klepfish, J.~B.~Kogut and M.~P.~Lombardo,
  Nucl.\ Phys.\ Proc.\ Suppl.\  {\bf 60} A (1998) 220.

\bibitem{Fodor:2001au}
  Z.~Fodor and S.~D.~Katz,
  Phys.\ Lett.\ B {\bf 534} (2002) 87.


\bibitem{FK1} 
 Z. Fodor and S. Katz, JHEP {\bf 0203} (2002) 014.

\bibitem{FK2}
 Z. Fodor and S. Katz, JHEP {\bf 0404} (2004) 050.

\bibitem{Philipsen1} 
  A.~Hart, M.~Laine and O.~Philipsen,
  Phys.\ Lett.\ B {\bf 505} (2001) 141. 
%
 P.~de Forcrand and O.~Philipsen,
  Nucl.\ Phys.\ B {\bf 642} (2002) 290;


\bibitem{Philipsen2} 
P.~de Forcrand and O.~Philipsen,
  Nucl.\ Phys.\ B {\bf 673} (2003) 170; 
%
O. Philipsen and Ph. de Forcrand, 
hep-lat/0409034.

\bibitem{ImagmuLombardo}
M.-P. Lombardo, Nucl. Phys. Proc. Suppl. {\bf 83} (2000) 375;
%
M. D'Elia, M.-P. Lombardo,
Phys. Rev. D {\bf 67} (2003) 014505;
%
M. D'Elia and M.P. Lombardo, 
Phys. Rev. D {\bf 70} (2004) 074509.

\bibitem{STV2}
K. Splittorff, D. Toublan, J.J.M. Verbaarschot, 
Nucl. Phys. B {\bf 639} (2002) 524. 

\bibitem{Kogut}
 J.B. Kogut and D.K. Sinclair, Phys. Rev. D {\bf 66} (2002) 034505 and
 Phys. Rev. D {\bf 70} (2004) 094501.

\bibitem{KTV}
  B.~Klein, D.~Toublan and J.~J.~M.~Verbaarschot,
  Phys.\ Rev.\ D {\bf 68} (2003) 014009.


\bibitem{HMZ}
L. He, M. Jin, P. Zhuang, 
hep-ph/0503272.

\bibitem{VJ}
B. Vanderheyden, A. D. Jackson, 
Phys. Rev. D {\bf 64} (2001) 074016.


\bibitem{Nishida}
Y. Nishida, 
Phys. Rev. D {\bf 69} (2004) 094501.

\bibitem{SLV}
  K.~Splittorff, J.~T.~Lenaghan and J.~Wirstam,
  Phys.\ Rev.\ D {\bf 67} (2003) 105011.

\bibitem{Shinji}
 S.~Ejiri,
 Phys.\ Rev.\ D {\bf 69} (2004) 094506.

\bibitem{O}  J. C. Osborn, Phys. Rev. Lett. {\bf 93} (2004) 222001.

\bibitem{AOSV}
G.~Akemann, J.~C.~Osborn, K.~Splittorff and J.~J.~M.~Verbaarschot,
Nucl. Phys. B {\bf 712} (2005) 287.

\bibitem{OSV}
 J.C. Osborn, K. Splittorff, J.J.M. Verbaarschot, hep-th/0501210, accepted
 for publication in Phys. Rev. Lett.

\bibitem{diracPQnum}  J.~B.~Kogut, H.~Matsuoka, M.~Stone, H.~W.~Wyld,
  S.~H.~Shenker, J.~Shigemitsu and D.~K.~Sinclair, 
  Nucl.\ Phys.\ B {\bf 225} (1983) 93;
I.~Barbour, N.~E.~Behilil, E.~Dagotto, F.~Karsch, A.~Moreo, M.~Stone and H.~W.~Wyld, 
  Nucl.\ Phys.\ B {\bf 275} (1986) 296;
M.~P.~Lombardo, J.~B.~Kogut and D.~K.~Sinclair,
  Phys.\ Rev.\ D {\bf 54} (1996) 2303;
S.Hands, I.Montvay, L.Scorzato, and J.Skullerud, 
Eur. Phys. J. C {\bf 22} (2001) 451.  

\bibitem{diracPQana} M. Stephanov, Phys. Rev. Lett. {\bf 76}, 4472 (1996); 
D. Toublan, J.J.M. Verbaarschot, Int. J. Mod. Phys. B {\bf 15} (2001) 1404; 
K. Splittorff and J.J.M. Verbaarschot, 
Nucl. Phys. B {\bf 683}, 467 (2004).



\end{thebibliography}
\end{document}